\documentclass[twoside]{article}

\usepackage{lipsum}
\usepackage{mathtools}
\usepackage{cuted}
\usepackage{rotating}
\usepackage{longtable}
\usepackage{pdflscape}
\usepackage[accepted]{aistats2022}
\begin{document}

\twocolumn[

\aistatstitle{Memory visualization tool for training neural network}

\aistatsauthor{ Mahendran N}

\aistatsaddress{ Indian Institute of Technology, Tirupati} ]

\begin{abstract}
  Software developed helps world a better place ranging from system software, open source, application software and so on. Software engineering does have neural network models applied to code suggestion, bug report summarizing and so on to demonstrate their effectiveness at a real SE task. Software and machine learning algorithms combine to make software give better solutions and understanding of environment. In software, there are both generalized applications which helps solve problems for entire world and also some specific applications which helps one particular community. To address the computational challenge in deep learning, many tools exploit hardware features such as multi-core CPUs and many-core GPUs to shorten the training time. Machine learning algorithms have a greater impact in the world but there is a considerable amount of memory utilization during the process. We propose a new tool for analysis of memory utilized for developing and training deep learning models. Our tool results in visual utilization of memory concurrently. Various parameters affecting the memory utilization are analysed while training. This tool helps in knowing better idea of processes or models which consumes more memory.
\end{abstract}

\section{INTRODUCTION}
Deep learning (DL) and other machine learning (ML) techniques are evolving in a rapid phase. Integration of machine learning algorithms with Software Engineering (SE) is latest trend. Reusable deep learning models are deployed while integrating with SE. Models once developed by the trained machine learning experts can be easily deployed  without the help of experts \cite{Li18}. Deep Learning advancements in Software Engineering, healthcare, Computer vision, Natural Language processing, Autonomous vehicles help enable remarkable progress in recent years. Processes like predicting sales, detection of disease using computer vision enhances idea of involvement of deep learning models in Software Engineering. Machine learning have a huger impact in Software Engineering as machine learning helps to generate code. 

Big tech companies are researching in the field of integrating AI with their SE tasks like code generation, malware detection. Integration of deep learning in SE tasks are becoming popular. Researches in this field consisted of integration more than 40 Software Engineering tasks are integrated with deep learning. There are research papers accepted in more than 60 venues about combining Deep learning with SE tasks. Machine learning algorithms learn by itself as it is not hand coded like Software. As machine learning algorithms are changing according to requirements and trends, the best practices in before deploying models in productions is to have exactness of the model.To know the quality of model before deploying in Software. To know the quality of model is important before deployment in order to have a good software model. The defective model has to be identified and with help of precision index \cite{Li18,Ma07}.
Although software benefit from deep learning, it comes with more utilization of memory than the normal software without deep learning algorithm. However, the models developed in the availability of large computational resources can be implemented in software easily. Cloud platforms resources paves the way for implementing even bigger machine learning algorithms in software effectively. But having offline models in software is reliable, secure. Memory utilization can cause training models not possible. There are several steps taken in order to reduce the memory utilization in deep learning algorithms.

Research going in this field to reduce memory consumption to an extent. Existing researches like compression of models, using cloud storage for ML models helps software to be applied for even less computational and less storage devices. Uncompressed networks like AlexNet (handles 150,528-dimensional input) and VGG (Very deep convolutional network containing 16-19 layers) have large numbers (more than tens of mega-bytes) of weights in fully connected layers which incur significant amount of memory accesses \cite{Krizhevsky12,Simonyan14}. And mobile devices have strict constraints in terms of computing power, battery, and memory capacity. \cite{Kim15} researched that compression of models results in reduction of memory, runtime and energy significantly with only minor accuracy drop. Compression can be performed after training the model in a environment where resources are available. In less computational environment there has to be steps taken in order to reduce the resource utilization like memory. Knowing where in the training model, memory is utilized more can help to solve this problem. The reason for increase/decrease in memory consumption at a particular time is not known. Storing models in cloud may be a better implementation when compared to compressing the models. This allow less memory usage in software applications but has issues like maintaining the storage, privacy of the model, internet connectivity, updating, versioning the model.

We have to find the root cause for increasing in memory utilization. Real time memory usage has to be known which paves the way to analyze each model while training. As large number of weights incur significant amount of memory access, there are certain other factors which increases the memory utilization. The increase in nodes for each layer, increase in layers, training epochs, computation complexity, input data dependency value or quality of image used in dataset contributes to increase in utilization of memory. Increase in memory utilization in turn increases the power consumption. Analysing the memory is needed for each models irrespective of hardware used. So having a tool for monitoring memory utilization in real time can provide better understanding of the current situation. We propose a tool which can monitor the memory utilization in computer and can produce the result in graph form.

These models have grown in memory requirements (typically in the millions of parameters or weights). These algorithms require high levels of computation during training as they have to be trained on large amounts of the data while during deployment they may be used multiple times. On the other side, these models do memory access each time when models are trained. So it is very important to consider memory usage while training a model.

\section{LITERATURE REVIEW}
The memory utilization plays an important role in developing a machine learning algorithm. Understanding the development of software helps in inheriting with the machine learning model. Software utilizes memory for storing the code and function it is intended to do. Software evolved in such a way, less memory utilized and can even inherited with wrist watches. Integrating software with machine learning also follows the same path so as to reduce the memory utilization to provide a better services.

\cite{marimuthu} et al. does empirical study on analysing energy in software applications. They found machine learning methods are adopted for detecting design pattern detection, code smell detection, bug prediction and recommending systems. Here the algorithms do have some patterns but when considering the energy consumption, only the small amount of energy is dissipated in these tasks. Where as the large amount of power dissipation is done in developing the model and when model undergoes training/testing phase. This paper provides idea of analysing the energy dissipation in machine learning models as similar to software applications.

\cite{Han15} et al. researched to reduce the storage and computation required by neural networks by an order of magnitude without affecting their accuracy by learning only the important connections. They does this to find energy and memory dissipation in neural networks. This can be achieved by changing weights and connections within neural networks. In our research, we measure the memory and energy spent on read/write operations by neural network or any machine learning algorithm. With the obtained results we get to know how much energy one particular algorithm is using and average of energy while training the model. 

\cite{Dayarathna15} et al. proposed a research survey on power consumption and memory consumption in Data centers. The data centers of the companies provide the space for most of the consumers data. Users data is critical in order to take any steps in growing the business for developing better product for users and so on. The survey details usage of consumption of energy in data centers. They analysed energy spent on cooling, lighting, power conversion, network and hardware and server and storage. They use multiple different power modeling tools for evaluation.

\cite{Kim15} et al. researched about finding the energy consumption for deep neural network from Titan x to Smartphones in order to compress the neural network.They are considering computation energy and memory access in deep learning. They try to compress the neural network so that they can fit it in smartphones. Caffe,Torch,Theano tools are used for developing neural networks. We provide results based on current analysis of energy dissipation in neural networks. Their tools and techniques which are used in extracting the energy dissipation in neural network.

\cite{Vanhoucke11} et al. researched to reduce the computation cost for training deep learning networks.They use specific modifications in order to improve the speed of neural networks by using floating point implementations. Our project is to find the energy spent on the neural networks.We are doing the calculation of computation and memory, where as they do some minor changes in order to increase the computational speed. Memory access data is also calculated. They did some implementations in memory access to reduce computation speed  but they did not actually considered the various types of neural networks. We can utilize the inspect their idea of calculating the computational speed for our machine learning algorithms such as neural networks.

\section{METHODOLOGY}
\begin{figure*}
    \centering
    \includegraphics[width=0.8\textwidth,height=10cm]{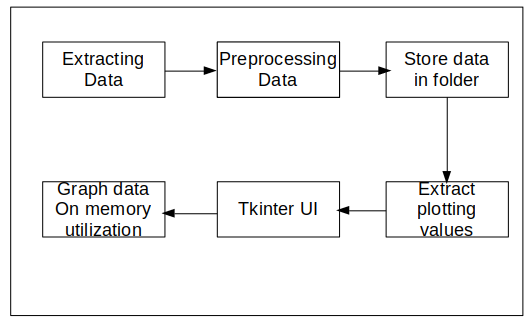}
    \caption{Overview of Tool development Approach}
    \label{fig:architecture}
\end{figure*}

This section introduces the details of the tool developed and its functions. We discuss the process of tool development, deployment of tool,execution of machine learning algorithm and management of concurrent extraction of measurements under this section. The core approach of how this tool works is depicted in Figure \ref{fig:architecture}

\subsection{Tool Development}
There needs to be a better user interface in order to get better user experience when using any tool for better understanding. The User Interface can be provided by TkinterUI in python. We develop a application type interface within the python package with the help of Tkinter. This package helps in providing user, the current status of the data and can update the memory readings in real time. 

\begin{figure}
    \centering
    \includegraphics[width=6cm,height=10cm]{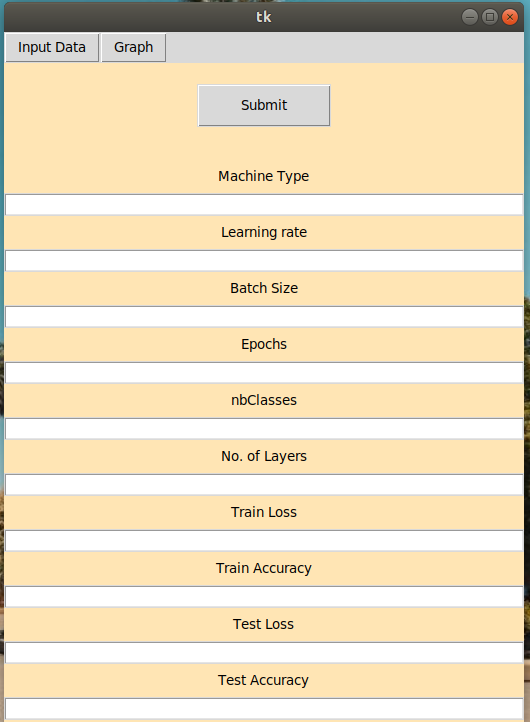}
    \caption{Tkinter User Interface}
    \label{fig:inputdata}
\end{figure}

The Tkinter User Interface has simple functions and needs less memory to implement in python. Figure \ref{fig:inputdata} depicts the Tkinter User Interface in python. In order to know the difference between the models trained, we provide the hyperparameter data input form in Tkinter as shown in Figure \ref{fig:inputdata}. With this data input we can identify the difference between the models trained and energy can be viewed in Graphical representation.


Tool development consists of taking crucial steps in order to get better outcome as tool. In the view of developing this tool, the basic requirements needed for this tool development are memory utilization values. The measurement of the memory utilization value has to be in real time. Real time values help in knowing the function of the model. Memory utilization is basically memory read and memory write signals. Multiple signals are needed will be high when performing numerous tasks at a time like matrix multiplications and so on. In order to get the memory utilization, we use 'psutil' directory in python. 

Psutil package helps in retrieving information on system utilization. We use Ubuntu environment for developing this project.The memory utilization in Ubuntu environment can be get using terminal commands like free,top and so on. Using the same commands the psutil package directly fetch the data and result it in seperate variable. We can call the variable whenever needed. It is much easier to use in the User Interface if memory utilization values are allotted to variable individually.

\subsection{Tool Deployment}
%
The execution of two functions and the tkinter UI can be performed by multiprocessing. In python, multiprocessing package helps in executing process simultaneously. This package supports spawning processes. Spawning function loads and executes a new child process. The current process and new child process can continue to execute concurrent computing. Here multiprocessing in python uses API similar to the threading module. The multiprocessing package offers concurrency on both locally and remote.

\begin{figure}
    \centering
    \includegraphics[width=6cm,height=10cm]{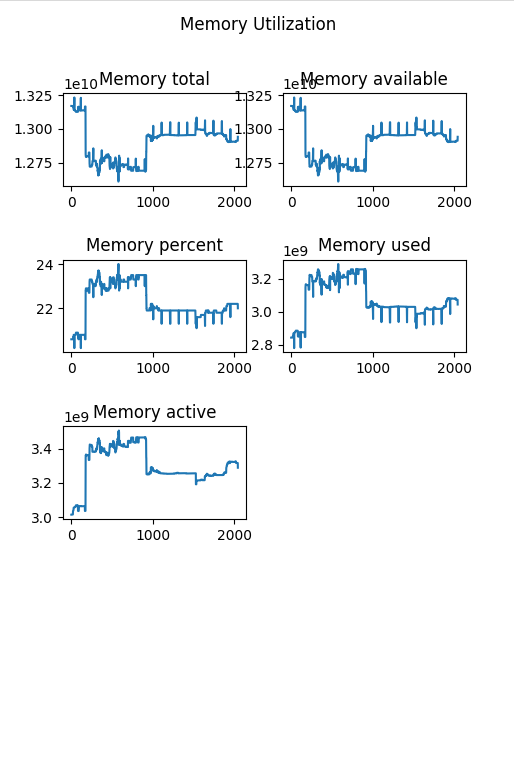}
    \caption{Memory utilization Graph}
    \label{fig:memorygraph}
\end{figure}

Once the tool is activated, the memory utilization data is extracted from the OS using the additional package 'psutil'. The extracted data is stored in the folder and concurrently the Tkinter UI reads the data from the stored database and provide a graphical format of memory utilization. The graphical form contains total memory, memory available, Memory used percentage, Memory available, memory currently active. Figure \ref{fig:memorygraph} represents the graphical data that is collected from the data stored.

\section{RESULTS}
We developed a tool for measurement of memory utilization and remembering the hyper parameters utilized while training a deep learning model. This helps in analyzing the memory utilization better for a deep learning model since knowing the hyperparameters and relate it with memory utilization. Utilizing this tool can help in developing better deep learning model keeping utilization of memory in consideration. 

\section{CONCLUSION}
We propose the new perspective of considering memory utilized while developing machine learning algorithms. Deep learning models are taken as example in providing an overview of tool and its functionalities.

\bibliography{sample_paper}

\end{document}